\documentclass[proof]{WileyASNA-v1}
\usepackage{siunitx}
\usepackage{natbib} 

\usepackage[utf8]{inputenc} 
\usepackage[T1]{fontenc}    
\usepackage{hyperref}       
\usepackage{url}            
 \articletype{Article Type}%

\received{..}
\revised{..}
\accepted{..}

\begin{document}

\title{The evolution of the UV/optical lag spectrum of NGC 7469 seen by the Liverpool Telescope}

\author[1,2,3]{F. M.  Vincentelli*}

\author[3]{M. Beard}

\author[3]{I. Mc Hardy }

\author[4]{E. Cackett}

\author[5]{K. Horne}

\author[6]{M. Pahari}

\authormark{ {VINCENTELLI ET AL}}

\address[1]{Instituto de Astrofisica de Canarias, E-38205 La Laguna, Tenerife, Spain}

\address[2]{Departamento de Astrofisica, Universidad de La Laguna, E-38206 La Laguna, Tenerife, Spain}

\address[3]{Department of Physics and Astronomy. University of Southampton, Southampton SO17 1BJ, UK}

\address[4]{Wayne State University, Department of Physics and\& Astronomy, 666 W Hancock St, Detroit, MI 48201, USA}

\address[5]{SUPA Physics and Astronomy, University of St. Andrews, North Haugh, KY16 9SS, UK}

\address[6]{Department of Physics, Indian Institute of Technology, Hyderabad 502285, India}
\corres{fvincentelli@iac.es}

\abstract{We present the results regarding the analysis of an intensive monitoring of the Active Galactic Nucleus (AGN) NGC 7469. We observed the source for 4 months with almost daily cadence in the ugriz bands, using the IO:O on the Liverpool Telescope. We measured the lags with respect to the u band and found a clear change of the lag spectrum between the first and the second half of the campaign. Given that the source varies on different timescales during these two segments, it is likely that different components are dominating the variability at different times. This result further confirms  that reverberation models require a more complex geometry than a static illuminating point source and that particular attention has to be given in the interpretation of these delays.  }


\maketitle

\section{Introduction}\label{sec1}

Active Galactic Nuclei (AGN) are believed to be powered by accretion onto supermassive black holes \citep{padovani2017,eht2019_bh}. Their multi-wavelength emission spans a wide range in the electromagnetic spectrum and is the sum of different physical components. It is generally accepted that the main engine of these systems is composed of an optically thick, geometrically thin disk  \citep[peaking in optical/UV,][]{shakura1973} and a hot Comptonizing region -- also known as corona--   in the innermost region of the system \citep[emitting in X-rays,][]{haardt_maraschi1991}.  However, despite being studied for decades, the exact geometry of their accretion flow, however, is still debated. 

Variability studies have revealed to be one of the most powerful tools to constrain the geometry of accreting objects \citep{uttley2014,cackett2021}. A clear example are the recent intensive monitoring campaigns lead by the \textit{Neil Gehrels Swift} Observatory with $\approx$ daily cadence, which aimed to map the structure of the accretion disk by studying the X-ray to UV/optical delays \cite{mchardy2014,mchardy2018,edelson2015,cackett2007,cackett2018,hernandez,vincentelli+2021,kara2021}. Interestingly, most of the sources showed delays inconsistent with the expectation from an irradiated accretion disk \citep[see e.g.][]{edelson2019}, motivating the development of models which required a radially or vertically extended corona \citep{kammoun2021,kammoun2021b,gardner2017}, an additional reprocessor \citep[as, for example, diffuse continuum from the broad line region][]{korista2001,korista2019,lawther2018,cheleouce2019} or propagating temperature fluctuations \citep{neust}.

Moreover, recent timescale and flux dependent studies are showing that, most probably, an interplay between these components is present in most of the sources \citep[see e.g.][]{cackett2018,cackett2021freq,mchardy2018,pahari2020,hernandez,vincentelli+2021,vincentelli+2022}.  However due to the small number of datasets and the \lq\lq limited" duration of these campaigns, it is impossible to draw general conclusions.

\begin{figure*}[h]
	\centering
	\includegraphics[width=2.1\columnwidth]{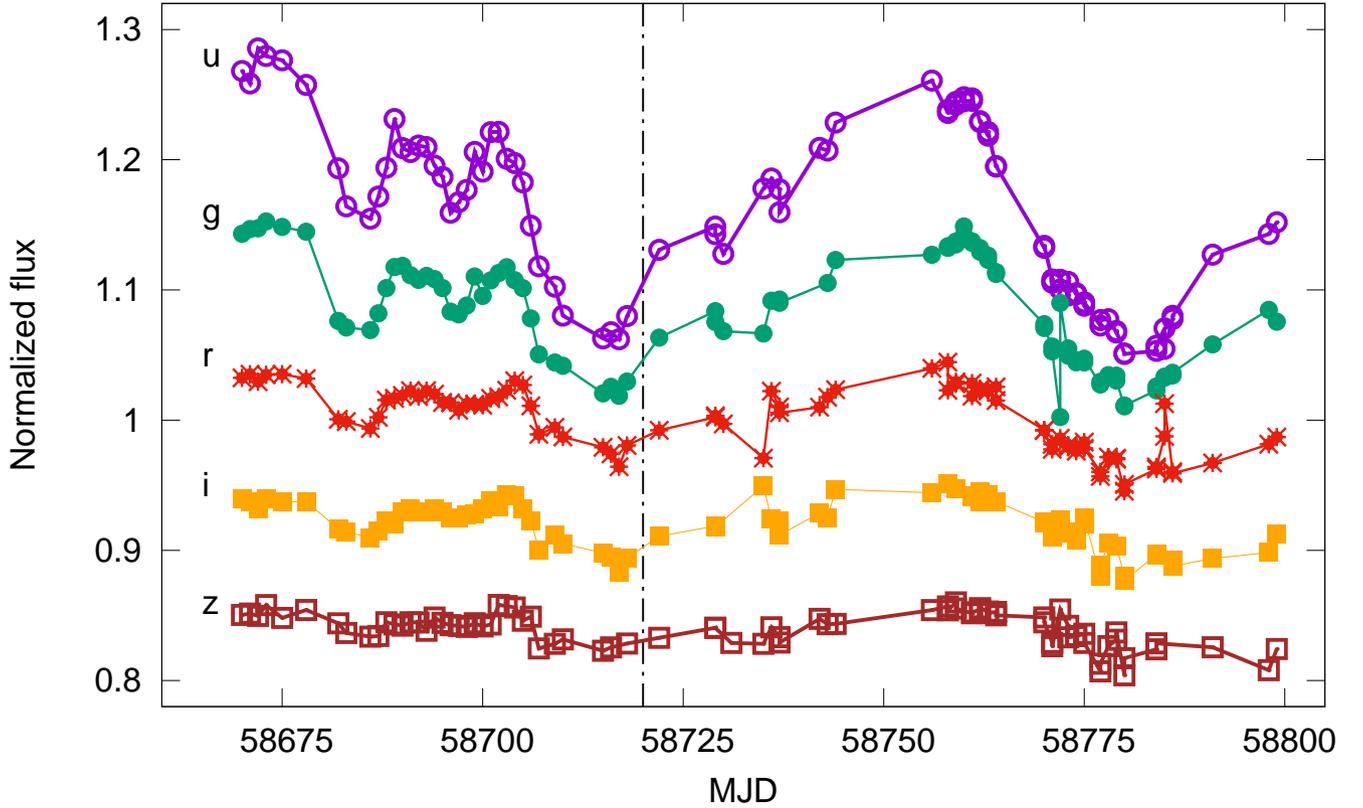}

    \caption{Normalized  lightcurves of NGC 7469 in the \textit{ugriz} bands with the Liverpool Telescope. Purple, green, red, orange and brown curves represent u (flux shifted of +0.16), g (shifted of +0.08), r, i (shifted of -0.08) and z (shifted of -0.16) band respectively. }
    \label{fig:lc}
\end{figure*}

 In this paper we present results regarding an intensive monitoring campaign on NGC 7469 \citep[$\approx 2 \times 10^7 M_\odot$][]{peterson2004}, one of the most studied AGN, well known for its strong variability in the UV-optical and X-ray bands. Early studies with RXTE and IUE  first showed that this object a presented a clear X-ray to UV lag of $\approx$ 4 days, which could not be explained with simple reprocessing \citep{nandra1998}. Following detailed analysis of this dataset showed that in order to explain such a delay, pure upscattering of UV photons was not enough, and requires also a variation in the coronal properties \citep{nandra2000,nandra2001,petrucci2001}

Recent analysis of this source by \citet{pahari2020}, showed that a lag-spectrum consistent with accretion disk reprocessing appears when filtering out the longer timescales variation. Motivated by these results we conducted a new optical continuum reverberation mapping campaign on NGC 7469. The paper is organized in the following way: Section 2 is focused on the data, Section 3 is focused on the analysis, while in section 4 we discuss the physical interpretation.

\section{Data}\label{sec2}

We observed NGC 7469 with the IO:O mounted on  the 2m Liverpool Telescope in  Roques de los Muchachos \citep{steele2004} between 6th of July and the 12 of November 2019  using the ugriz band (program PL19B15, PI: Vincentelli). We collected 90 observations which allowed us to get an average sampling of 1 observation every 1.4 days. The data was reduced using HiPERCAM's reduction pipeline \footnote{ \url{https://cygnus.astro.warwick.ac.uk/phsaap/hipercam/docs/html/}}. This used 4.2 arcsecond radius apertures to retrieve flux values for the object and a nearby non-variable comparison star. To filter out atmospheric variations, the lightcurve was retrieved as the relative flux between the AGN and the comparison star. As shown in Fig.\ref{fig:lc} the lightcurve is extremely well sampled and shows clear variability in all the bands down to $\approx$ few days timescales.

\begin{figure}
	\includegraphics[width=\columnwidth]{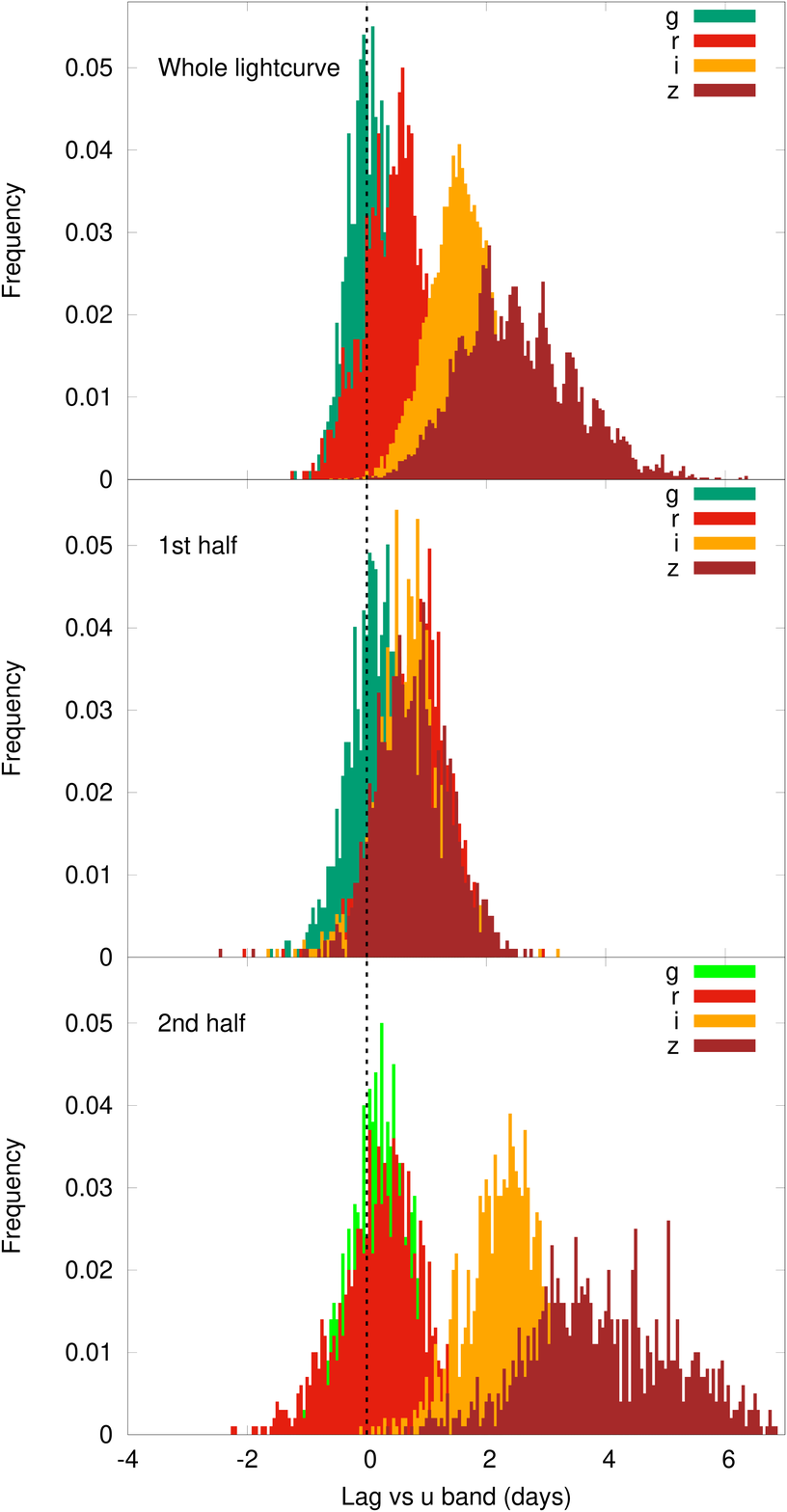}

    \caption{Distribution of the lag versus the u band centroids using the whole lightcurve (Top Panel), first half (central panel) and second half (bottom panel). The color code is the same as in Fig. 1. A clear evolution in the i and z band can be clearly seen.}
    \label{fig:example_figure}
\end{figure}

\section{Analysis}\label{sec3}

\subsection{Interband lags}
 We  computed the lags of the griz bands with respect to the u band. In particular, we used the method known as  “flux randomization (FR) and random subset selection (RSS)" \citep{peterson1998,peterson2004}. This method calculates the  cross correlation function (CCF) by re-sampling and randomizing the  fluxes at different epochs N(=10$^4$) times. The lag  between the lightcurves and its error is then evaluated from distribution of the centroid of the N CCFs. In Fig. \ref{fig:example_figure} (Top panel)  we show the resulting lag distribution of the whole lightcurve, which shows a clear evolution of the lags as a function of wavelength. The lags are  reported in Table 1.

\subsection{Timescale dependent analysis}

From a visual inspection of the lightcurve it is clear that  the first half shows variability patterns faster than the second half. {In the first half of the lightcurve, we can see variations on timescales of few days (i.e. it is possible to distinguish at least three clear peaks), while in the second half the lightcurve seem to show a smoother trend, without short timescales flaring (i.e. one single slower peak). Therefore, even though the data does not allow us to apply Fourier techniques\citep{cackett2021freq}, this behaviour gives us the rare opportunity  to explore the evolution of the lags for different timescales without applying any filtering to the lightcurve. Therefore, we divided the lightcurve in two and recalculated the lags for the two segments: before and after MJD 58720. This choice was done for two different reasons. First, it is well known that long term trends can strongly affect the lag evaluation \citep{white1994}. Thus the minimum around 58720 is an ideal position to separate the timescales without introducing any bias. Second, such a choice allowed us to keep a relatively high number of point for both segments ($\approx40$). Given also that the lags are not highly significant using the whole range ($\approx2-3\sigma$ in the i-z bands), a finer cut would not only further degrade the results and but also potentially introduce biases.   Yet we note that shifting the separation threshold of $\pm$10 days, the measured properties remained unchanged. }

\begin{figure}
	\includegraphics[width=\columnwidth]{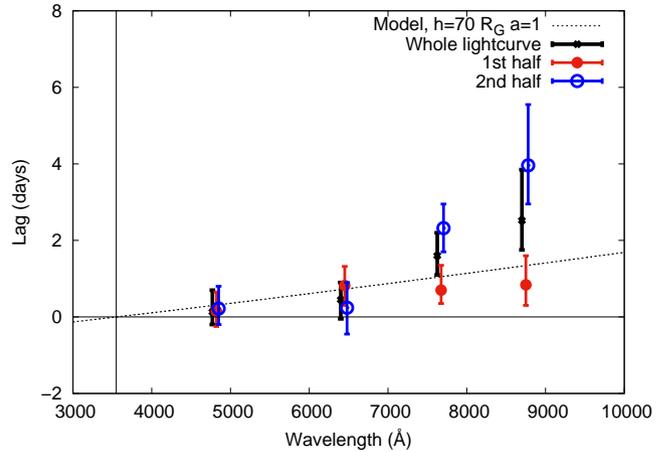}

    \caption{Lag spectra for the three sets of lightcurves (Grey points, whole lightcurve; red, first half; blue second half). The vertical line represents the reference band (u band), while the dashed line is the accretion disk reprocessing prediction fitted in the \citet{pahari2020} dataset by \citet{kammoun2021}.}
    \label{fig:lagspec}
\end{figure}

 The resulting lag distribution and lag spectra are shown in Fig. \ref{fig:example_figure} (central and bottom panel) and Fig. \ref{fig:lagspec} respectively. The lags are again reported in Table 1.

For the first half (short timescales, central panel in Fig. \ref{fig:example_figure}) all lags are close to $\approx$1 day, and we do not see evidence of an evolution as a function of wavelength, most probably due to large errors. This seems to be confirmed by  the inset of Fig. \ref{fig:lc} where all bands seem to follow the same pattern, almost simultaneously.

In the second half (longer timescales, bottom panel Fig. \ref{fig:example_figure}), the lower frequency part of the lag spectrum ( i and z) clearly evolves, shifting to longer lags, while the g and the r band do not change significantly. In the second half of the lightcurve, it is possible to see that all bands show a rise of $\approx$ 30 days and a decay of $\approx$ 20 days. As already seen in other sources \citep[see e.g. Mrk 110,][]{vincentelli+2021}, the response during the decay is slower at longer wavelengths and this introduces a longer lag.

\section{Discussion}\label{sec4}

The sample of AGN mapped with intensive photometric campaigns has rapidly increased in the last few years.  All studies seem to agree that the overall phenomenology cannot be explained with an accretion disk illuminated by a single point source \citep[see][and references therein]{cackett2021}. This lead to the development of new reverberation models, which seem to show the need for a vertically  or radially extended corona. Further studies, however, also indicated that the lag spectrum of these objects has a significant contribution from an additional reprocessor, usually associated with the Broad Line Region (BLR) \citep{mchardy2014,cackett2018,vincentelli+2021,vincentelli+2022}

Our monitoring of NGC 7469 further demonstrates this, showing that the lag spectrum is timescale dependent and can change significantly in time within a few months. Several studies have  demonstrated that this source requires a dynamical corona in order to reproduce the UV/X-ray negative lags of $\approx$4 days \cite{nandra2000,nandra2001,petrucci2001}. An expanding corona (vertically or radially) is expected modify the observed lag spectrum \citep{kammoun2021,kammoun2021b}. Thus, despite the lack of X-ray coverage, we attempted to quantify the properties of the accretion flow by using the equation 1 in \citet{kammoun2021b} for scaling of the lags as a function of the AGN property. In detail, we used the X-ray luminosity reported by \citet{pahari2020} \citep[long term monitoring did show a drastic evolution in X-ray flux][]{nandra1998,middeingc,pahari2020}. 

Interestingly, the lag spectrum from the first half of the campaign seems to require a spinning black hole, and is consistent with the parameters found by  \citet{kammoun2021b}.  However, the lags observed in the second half are too high to be fitted  {only} with this model, {even assuming a change in the coronal height}. In particular the apparent change of slope in the lag spectrum, suggests the presence of an additional component dominating the emission at longer wavelengths, on longer timescales.

{\citet{lawther2018} showed that the  response of the diffuse continuum from the BLR is expected respond better to longer timescales variations. Therefore, this component could be a potential candidate to explain the observed delays. As mentioned above, a very similar behaviour has already been observed in the Narrow line Seyfert galaxy  Mrk 110, with a  higher lag at longer wavelengths during a slower variation was already observed by \citet{vincentelli+2021}. This was interpreted as the slower decay timescale of the diffuse continuum emission\citep{korista2001,korista2019} from the BLR after a sudden increase of the accretion rate.  The lack of X-ray coverage for our campaign does not allow us to make a one to one comparison, but already a  change in the BLR properties due to the timescale of the driving signal could explain the observed evolution of the lag spectrum.}

We note that this does not exclude a possible evolution of the corona during the campaign.   A dynamically evolving corona,{ as the one observed with pure X-ray spectral timing methods\citep[e.g.][]{alston2020},} could play a significant role in at least driving the changing of the contribution from the diffuse continuum from the BLR to the overall multi-wavelength variability\citep{vincentelli+2021,vincentelli+2022}. {Although, details models which take into account both disk geometry and BLR contribution are required\citep[see e.g. ][]{jaiswal2022}, further observations could also help to disentangle between these two different phenomena. For example, by performing multi-epoch detailed X-ray spectral- timing analysis at different stages of the campaign could help to constrain the inner geometry of the accretion flow\citep{mastroserio2020} and therefore quantify the real contribution of the BLR.}

\begin{table*}[]
\centering
\begin{tabular}{cccccccc}
 \hline

 Band &Wavelength (\si{\angstrom})  &   & Lags Whole LC (days) &  & Lags 1$^{st}$ half  (days)  &  & Lags 2$^{nd}$ half  (days)  \\
 \hline
 \\
g &4770& &0.13$_{-0.33}^{+0.57}$ & &0.15$_{-0.40}^{+0.50}$ &  & 0.22$_{-0.42}^{+0.58}$   \\
r &6400&  &0.45$_{-0.50}^{+0.45}$ & & 0.80$_{-0.52}^{+0.50}$ & & 0.24$_{-0.69}^{+0.66}$\\
i &7625&  &1.60$_{-0.50}^{+0.60}$ & & 0.70$_{-0.35}^{+0.65}$  & & 2.32$_{-0.62}^{+0.63}$\\
z &8700&  &2.52$_{-0.77}^{+1.33}$   & & 0.84$_{-0.54}^{+0.76}$  & & 3.96$_{-1.01}^{+1.59}$\\
\\
\hline
\end{tabular}
\caption{Resulting lag with respect to the u band for the different bands and the different set of lightcurves.}
\end{table*}

\section{Conclusions}\label{sec5}

We presented the results regarding the analysis of the interband lags of the highly variable AGN NGC 7469. Our measurements, in line with previous observations of other AGN, indicate that the UV to optical lags are strongly dependent on the the timescale of the variability, suggesting a significant contribution from the BLR. These results, therefore, clearly show the need for further and longer multi-wavelength campaigns on these objects in order to adequately interpret the role of the different components and their complex phenomenology.

\bibliography{Wiley-ASNA.bib}%

\section*{Acknowledgments}

{The authors are grateful to the anonymous referee for the comments which significantly improved the manuscript. This project was supported by the STFC funding ST/R000638/1. FMV acknowledges financial support from grant FJC2020-043334-I financed by MCIN/AEI/10.13039/501100011033 and NextGenerationEU/PRTR.
} 

\end{document}